\documentclass[a4paper,11pt]{article}
\usepackage{jheppub}

\usepackage{amssymb,amsmath,amsfonts,bbm}
\usepackage{tikz}

\newcommand{\be}{\begin{equation}}
\newcommand{\ee}{\end{equation}}
\newcommand{\bea}{\begin{eqnarray}}
\newcommand{\eea}{\end{eqnarray}}
\newcommand{\nn}{\nonumber}

\def\Z{ {\mathbb Z} }
\def\R{ {\mathbb R} }

\begin{document}
\begin{flushright}
HIP-2016-13/TH
\end{flushright}

\title{\boldmath Holographic anyonization: A systematic approach}

\author[a]{Matthias Ihl,}
\author[b,c]{Niko Jokela,}
\author[b,c]{and Tobias Zingg}
\affiliation[a]{Centro de F\'isica do Porto e Departamento de F\'isica e Astronomia\\
Faculdade de Ci{\^e}ncias da Universidade do Porto\\ Rua do Campo Alegre 687, 4169-007 Porto, Portugal}
\affiliation[b]{Department of Physics and $\;^{\mathrm{c}}$Helsinki Institute of Physics\\
P.O.Box 64, FIN-00014 University of Helsinki, Finland}

\emailAdd{matthias.ihl@fc.up.pt}
\emailAdd{niko.jokela@helsinki.fi}
\emailAdd{tobias.zingg@helsinki.fi}

\abstract{Anyons have garnered substantial interest theoretically as well as experimentally.
Due to the intricate nature of their interactions, however, even basic notions such as the equation of state for any kind of anyon gas have eluded a profound understanding so far.
Using holography as a guiding principle, we propose a general method for an alternative quantization of electromagnetic degrees of freedom in the gravitational dual to obtain an effective physical description of strongly correlated anyonic systems. We then demonstrate the application of this prescription in a toy model of an anyonic fluid at finite charge density and magnetic field, dual to a dyonic black brane in $AdS_4$, and compute the equation of state and various transport coefficients explicitly.}

\vspace{6mm}
\numberwithin{equation}{section}
\setcounter{footnote}{0}
\renewcommand\thefootnote{\mbox{\arabic{footnote}}}

\maketitle
\flushbottom

\section{Introduction}\label{sec:intro}
The theoretical concept of anyons emerged in the late 1970s~\cite{Leinaas:1977fm}, based on the observation that the spin statistics theorem is less restrictive in (2+1) dimensions and allows for types of particles with 'any' value of spin.
The resulting anyons have a multitude of novel properties that are still subject to intensive research -- see, \emph{e.g.},~\cite{Rao:1992aj,Stern:2008} for reviews.
Probably the most striking feature is that they obey fractional statistics, {\emph{i.e.}}, a statistics that interpolates between the Fermi--Dirac and the Bose--Einstein distribution, the interpolation parameter often being referred to as statistical angle.
A major surge in the investigation of anyons came after the discovery that the fractional quantum Hall effect (FQHE) has a natural explanation using abelian anyons and, subsequently, when it became clear that an anyon gas coupled to a dynamic electromagnetic field could become superconducting~\cite{Chen:1989xs}.
Furthermore, it has been speculated that non-abelian anyons can also be realized in the FQHE~\cite{Moore:1991ks}, which makes them a very interesting topic to investigate and apply in topological quantum computing~\cite{Kitaev:1997wr}.

Albeit some progress in the understanding of many anyon systems in the high temperature, low density (virial expansion of the equation of state) and low temperature, high density (mean field approach) limits has been made, not much is known about anyons beyond the two anyon case in weakly-coupled field theory.
For example, owing to the fact that multi particle anyon states cannot be expressed as a simple product of single particle states, the exact calculation of the grand partition function of an anyon gas is still elusive.
In the same vein, the virial expansion of the equation of state for anyons is only known up to the second virial coefficient -- as the latter only depends on two-body interactions in a quantum cluster expansion.

With this in mind, it seems important to take a step back and ask whether and in what regimes one can reasonably expect perturbative methods to be applicable.
In fact, there are simple arguments indicating that the mere non-trivial exchange statistics can be viewed as hidden Chern--Simons type interactions~\cite{Hosotani:1992jm}. Model calculations show that the inclusion of even modest long-range repulsive interactions have drastic, orders of magnitude, effects on spontaneous magnetization~\cite{YiCanright}.
It is thus not really surprising that the free energy of a collection of anyons does not resemble the sum of energies of free anyons. This in turn justifies the attempt to conceive an altogether different description of a many-anyon systems at strong coupling.

In this work, we will make use of the gauge/gravity, or holographic, duality, see~\cite{AdS_CFT_reviews} for reviews, to address
anyons at strong coupling. 
Of course, the strongly-coupled field theory duals are quite different from the usual weakly-coupled field theories which one normally encounters in condensed matter physics. In light of the limited progress made concerning weakly-coupled anyons, it seems plausible that their holographic cousins may help to shed some light on this complicated subject.
The subject of this work are four-dimensional bulk theories whose boundaries have a dual description as (2+1)-dimensional gauge field theories.
We are interested mainly in gauge excitations and thermodynamics of these boundary field theories in order to, ultimately, make contact with the description of anyonic fluids.

According to holography, the charge density and magnetic field of  the matter fields are encoded in the boundary data of the bulk gauge fields. The standard choice is to pick boundary conditions for these gauge fields such that either chemical potential or charge density is fixed, depending on the thermodynamic ensemble, but most importantly with fixed magnetic field strength. This precludes the gauge field from being dynamical.
An equally natural choice would be to consider Robin boundary conditions, \emph{i.e.}, an interpolation between Neumann and Dirichlet conditions~\cite{Witten:2003ya,Yee:2004ju}.
This, in particular, allows the magnetic field to become dynamical and thus letting it adjust its own vacuum expectation value.
Such boundary conditions lead to mixing electric and magnetic charges on the boundary, resulting in rendering the charge carriers anyonic.
Building upon the prescription pioneered in~\cite{Jokela:2013hta}, our proposal invokes a different procedure to anyonize a given system.
This is a complementary framework, but has the benefit that it provides a consistent method to deal with residual gauge degrees of freedom that would lead to ambiguities on the boundary field theory.
This is the topic of section \ref{sec:anyonizing}.

We then proceed in section~\ref{sec:application} with a discussion about how thermodynamic quantities and transport properties are affected by applying the aforementioned $SL(2,\mathbb{Z})$ mapping.
As an accompanying illustrative example, we demonstrate this application of our procedure to a dyonic Reissner--Nordstr\"om black brane in $AdS_4$.
This solution to the equations of motion resulting from the Einstein--Maxwell action is well-known as a dual model to holographic matter at finite charge density and magnetic field.
Furthermore, it allows us to explicitly compute the transformed grand potential and transport coefficients, as well as provide a consistency check of all transformation laws for our anyonization procedure.
We also derive the equation of state and expand it at high temperatures, where the analogy with the virial expansion is most transparent. 
In the final section~\ref{sec:discussion}, we summarize the results and discuss possible future directions and open problems.
Appendix~\ref{app:var} provides more details on the variation of the boundary action.


\section{Holographic anyons}
\label{sec:anyonizing}

In this section we will outline a general method for anyonizing holographic matter.
Firstly, a few introductory comments are provided on previous work including a discussion of some shortcomings related to ambiguities associated with gauge freedom.
We then proceed to propose a different prescription which will ultimately manage to circumvent such ambiguities and describe how the two prescriptions are connected. To demonstrate the versatility of this novel method, the transformed Green's functions are obtained and computed, along with further thermodynamic quantities and transport properties.

\subsection{Alternative quantization}

The procedure of alternative quantization\footnote{For a self-contained review in the present context, see~\cite{Jokela:2013hta}.} transforms a (2+1)-dimensional conformal field theory (CFT) into another by changing the boundary conditions on the bulk gauge field. In modern language, this procedure is an $SL(2,\mathbb{Z})$ electromagnetic transformation.
Therefore, as the name suggests, the manner in which boundary degrees of freedom are separated into source and response is changed, while the bulk description and equations of motion remain untouched.
It is important to note that the bulk action does not need to be invariant under $SL(2,\mathbb{Z})$,
we only demand the bulk gauge equations of motion reduce to free ones close to the boundary where Robin, \emph{i.e.}, combined Dirichlet/Neumann, boundary conditions are imposed.\footnote{This should be contrasted with~\cite{Goldstein:2010aw,Bayntun:2010nx,Lippert:2014jma,Fujita:2012fp}, where the $SL(2, \mathbbm{Z})$-symmetry was imposed in the bulk as well.}
In fact, one could allow for a larger group of transformations, but keeping in mind a possible, and desirable, embedding into string theory, where both the electric charges and magnetic fields are integer-quantized, one restricts to $SL(2,\mathbb{Z})$.

To be more specific, an anyon can be viewed as a quasiparticle consisting of a boson or fermion with some additional fixed amount of magnetic flux attached per fundamental unit of charge. In a conformal field theory with a global $U(1)$ one can perform the process of adding magnetic fluxes using an $SL(2,\mathbb{Z})$ electromagnetic transformation~\cite{Witten:2003ya, Yee:2004ju}. Under this transformation, the original CFT maps into another one with mixed charges and magnetic fields and thus the charge carriers have been transformed into anyonic degrees of freedom. From the bulk gravitational point of view, one is choosing an alternative quantization scheme for the gauge fields, \emph{i.e.}, the boundary values on the gauge fields have combined Dirichlet/Neumann conditions. This was first implemented holographically in~\cite{Jokela:2013hta} by showing that the $SL(2,\mathbb{Z})$ transformation on a fractional quantum Hall state~\cite{Bergman:2010gm} leads to a soft mode which is a prerequisite of an anyonic superfluid. Successful extensions to flowing superfluids~\cite{Jokela:2014wsa}, to other D-brane models~\cite{Brattan:2013wya,Brattan:2014moa,Itsios:2015kja}, and even extensions to backgrounds not dual to CFTs~\cite{Jokela:2015aha,Itsios:2016ffv} have been constructed subsequently. 

However, the situation is still incomplete and somewhat unsatisfactory. In the above-mentioned works, the main focus has been on the fluctuation spectra of the corresponding anyon fluids and on how these are affected by the 
choice of combined Dirichlet/Neumann boundary conditions. The addition of boundary terms to the bulk action is enough to implement the alternative quantization for the gauge fields. However, there is an unfixed gauge freedom left, an issue which needs to be addressed in the computations of the free energy.

\subsection{Helmholtzian prescription}
Our primary goal is to revisit the alternative quantization scheme for electromagnetism in a more formal and rigorous fashion.
We treat the electric and magnetic degrees of freedom on equal footing and introduce an auxiliary potential for the magnetic field.
It is explained why the Helmholtz potential will remain invariant under linear transformation of the currents and shown that the variation of the action will lead to the same expression as in previous works, but in a way that does not leave any ambiguities in the transformed fields.

We consider a generic action on a manifold $M$, with a Lagrangian density $\mathcal{L}[A,dA]$ which is invariant under $U(1)$ gauge transformations of $A$.
Using $\star$ to denote the Hodge dual on the (2+1)-dimensional boundary $\partial M$, an on-shell variation of the action with respect to all boundary degrees of freedom reads,\footnote{For a detailed discussion and derivation of the on-shell variation, see app.~\ref{app:var}.} 
\bea
\delta S
	&=&	\int_{\partial M} \delta A\wedge \star \mathcal{J}
		+ \int_{\partial M} \delta \mathcal{B} \wedge \star \eta
\, . \label{eq:delta_S_onshell}
\eea
Here, $\mathcal{J}$ denotes the electric current and $\mathcal{B} = - \frac{1}{2 \pi} \star^{-1} dA\bigr|_{\partial M}$ the current related to the magnetic flux through $\partial M$.
This interpretation can be illustrated by considering,
\be
\Phi = -\int_{t=const \atop r=const} \star\mathcal{B} \, ,
\ee
which is the magnetic flux through a surface element at time $t$, which is actually independent of $r$, \emph{i.e.}, constant along the radial flow.
The magnetic potential $\eta$ is essentially defined through \eqref{eq:delta_S_onshell} and more details about how it is related to the bulk fields can be found in the appendix.
However, for all practical purposes, namely for the geometries of interest in what follows, the only relevant component is the one from radial integration,
\be
\eta = - 2 \pi \int_{r_{H}}^{r} \frac{\partial\mathcal{L}}{\partial\, dA}
\, ,
\ee
where $r$ denotes a radial coordinate and $r_{H}$ is, depending on the bulk configuration, the position of the event horizon or the origin.

One possibility to perform an alternative quantization is to go from the original action to the 'Helmholtzian' form by adding a boundary term and making a Legendre transformation in $A$,
\bea
S_{{a}}[\mathcal{J},\mathcal{B}]
	&:=&	S - \int_{\partial M} A \wedge \star \mathcal{J} \, . \label{eq:S_Helmholtz}
\eea
This exchanges the role of $A$ and its canonical momentum and can also be interpreted as going from Dirichlet to Neumann boundary conditions. 
Furthermore, by construction,
\bea
\delta S_{{a}}
	&=&	- \int_{\partial M} \delta \mathcal{J} \wedge \star A
		+ \int_{\partial M} \delta \mathcal{B} \wedge \star \eta \, . \label{eq:delta_S_Helmholtz}
\eea
The Helmholtz action~\eqref{eq:S_Helmholtz} will turn out to be useful for our purposes, since it is an explicit functional of the electric and magnetic current, which parameterize the degrees of freedom of the field strength $F$. As such, they do not depend on an explicit choice of gauge and the transformations that follow can, if needed, be defined off-shell as well.
Moreover, these degrees of freedom can also be identified with normal and parallel directions relative to the boundary.
Quite generally, the bulk Lagrangian $\mathcal{L}$ is a functional of $|F|^2$, at least in the asymptotic region; in this case it is straightforward to use (\ref{eq:J}) to conclude $\mathcal{J} \propto \imath_n F$, \emph{i.e.}, the current contains only the components orthogonal to $\partial M$.
By construction, $\mathcal{B}$ depends only on parallel directions.
On-shell, a mixing of these currents, or respectively the corresponding degrees of freedom, would result in a mixing of Dirichlet and Neumann boundary conditions of the gauge field $A$, as described in the formulation used in~\cite{Jokela:2013hta}.

Using the Helmholtz form with $\mathcal{J}$ and $\mathcal{B}$ as dependent variables may, at first sight, seem more abstract, but it provides a remarkably simple and straightforward way to evaluate the action in an alternative ensemble as a linear transformation of variables.

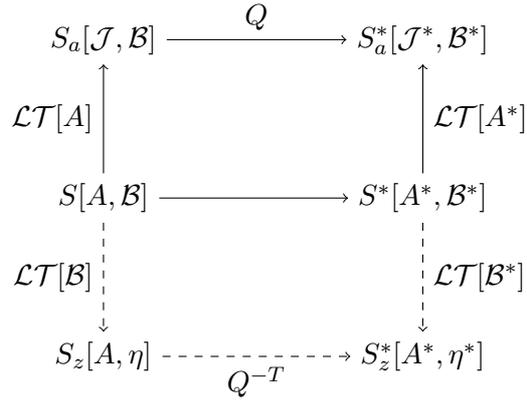
\begin{figure}
\begin{center}
\begin{tikzpicture}[node distance=21mm, auto]
  \node (S)  {$S[A,\mathcal{B}]$};
  \node (Sa) [above of=S]{$S_{{a}}[\mathcal{J},\mathcal{B}]$};
  \node (Sz) [below of=S] {$S_{{z}}[A,\eta]$};
  \node (dummy) [right of=S]{};
  \node (Sstar) [right of=dummy]{$S^{\ast}[A^{\ast},\mathcal{B}^{\ast}]$};
  \node (Sastar) [above of=Sstar]{$S_{{a}}^{\ast}[\mathcal{J}^{\ast},\mathcal{B}^{\ast}]$};
  \node (Szstar) [below of=Sstar] {$S_{{z}}^{\ast}[A^{\ast},\eta^{\ast}]$};
  \draw[->] (S) to node [above] {} (Sstar);
  \draw[->] (S) to node [left] {$\mathcal{LT}[A]$} (Sa);
  \draw[->] (Sstar) to node [right] {$\mathcal{LT}[A^{\ast}]$} (Sastar);
  \draw[->] (Sa) to node [above] {$Q$} (Sastar);
  \draw[->, dashed] (S) to node [left] {$\mathcal{LT}[\mathcal{B}]$} (Sz);
  \draw[->, dashed] (Sstar) to node [right] {$\mathcal{LT}[\mathcal{B}^{\ast}]$} (Szstar);
  \draw[->, dashed] (Sz) to node [below] {$Q^{-T}$} (Szstar);
\end{tikzpicture}
\end{center}
\caption{Schematic diagram of how the action $S[A,\mathcal{B}]$ is related to the 'alternative' action $S^{\ast}[A^{\ast},\mathcal{B}^{\ast}]$ .
Each of them can be brought to the Helmholtz form via a Legendre transform ($\mathcal{LT}$) in the dependent variable $A$, respectively $A^\ast$.
The two resulting actions, $S_{{a}}[\mathcal{J},\mathcal{B}]$ and $S_{{a}}^{\ast}[\mathcal{J}^{\ast},\mathcal{B}^{\ast}]$, are related via the linear transformation of variables  described in \eqref{eq:anyon-trafo}.
Alternatively, a path could have been chosen where a Legendre transform is performed in the dependent variables $\mathcal{B}$ and $\mathcal{B}^\ast$, respectively.
All directions commute.}
\label{fig:actiondiagram}
\end{figure}

Consider the transformation,
\be
\left(
\begin{array}{cc}
	\mathcal{J}^{\ast}	\\
	\mathcal{B}^{\ast}
\end{array} \right)
= Q \cdot \left(
\begin{array}{cc}
	\mathcal{J}	\\
	\mathcal{B}
\end{array} \right)	\, , \quad
\left(
\begin{array}{cc}
	A^{\ast}	\\
	-\eta^{\ast}
\end{array} \right)
= Q^{-T} \cdot \left(
\begin{array}{cc}
	A	\\
	-\eta
\end{array} \right)	\, , \quad
Q = \left[
\begin{array}{cc}
	a_s	& b_s	\\
	c_s	& d_s  \end{array} \right] \, .
\label{eq:anyon-trafo}
\ee
For the time being, we consider $Q$ as a linear transformation between vector spaces -- in section~\ref{sec:GF} we will explain why we restrict to $SL(2,\mathbb{Z})$ in the following.
Then, define the Helmholtz form of the action with alternative boundary conditions,
\bea
S_{{a}}^{\ast}[\mathcal{J}^{\ast},	\mathcal{B}^{\ast}]
	&:=&	S_{{a}}[Q^{-1}(\mathcal{J}^{\ast},	\mathcal{B}^{\ast})] \, .
\label{eq:S_anyon_Helmholtz}
\eea
This is nothing more than a linear change of variables, and the differential can be straightforwardly evaluated via a direct application of the chain rule,
\bea
\delta S_{{a}}^{\ast}
	&=&	- \int_{\partial M} \delta \mathcal{J}^{\ast} \wedge \star A^{\ast}
		+ \int_{\partial M} \delta \mathcal{B}^{\ast} \wedge \star \eta^{\ast} \, . \label{eq:delta_S_Helmholtz_star}
\eea
As a final step, this action can be Legendre-transformed once more in order to arrive at an expression that can be interpreted as the action related to the grand-canonical potential in an alternative quantization scheme,
\be
\boxed{S^{\ast}
	:=	S_{{a}}^{\ast} + \int_{\partial M} A^{\ast} \wedge \star \mathcal{J}^{\ast}
	=	S - \int_{\partial M} A \wedge \star \mathcal{J} + \int_{\partial M} A^{\ast} \wedge \star \mathcal{J}^{\ast}}
\, . \qquad\label{eq:S_star}
\ee
This is one of the main results in this paper. Moreover, the variation is easily evaluated,
\bea
\delta S^{\ast}
	&=& \int_{\partial M} \delta A^{\ast} \wedge \star \mathcal{J}^{\ast}
		+ \int_{\partial M} \delta \mathcal{B}^{\ast} \wedge \star \eta^{\ast} \, . \label{eq:delta_S_star}
\eea
To demonstrate the consistency of how $S^\ast$ was constructed, consider a different Legendre transformation of the original action,
\bea
S_{{z}} = S_{{z}}[A,\eta]
	&:=&	S - \int_{\partial M} \eta \wedge \star \mathcal{B}  \, . \label{eq:S_antiHelmholtz}
\eea
This is to be treated as a functional of the potentials $A$ and $\eta$, and thus allows an equally convenient way to perform the transformation \eqref{eq:anyon-trafo}. 
By proceeding in an analogous manner as before, this allows to take a different route of Legendre transformations and field redefinitions and to arrive at a second expression for the action related to the grand potential in the alternative ensemble,
\bea
\widetilde{S}^{\ast}
	&:=&	S_{{z}}^{\ast} + \int_{\partial M} \eta^{\ast} \wedge \star \mathcal{B}^{\ast}  \, . \label{eq:S_star_alt}
\eea
Comparing this to \eqref{eq:S_star}, and taking the relations \eqref{eq:anyon-trafo} into account, it can easily be verified that the difference ${S}^{\ast}-\widetilde{S}^{\ast}$ vanishes identically, showing that both expressions are indeed equal.
We illustrate the different routes of applying the alternative quantization in a commutative diagram in figure~\ref{fig:actiondiagram}.

\subsection{Anyonizing and Green's functions}
\label{sec:GF}

Now, let us turn towards the meaning of the transformed boundary conditions for the dual field theory.
With regard to anyons, there are two transformations of changing one CFT into another one that are of main interest. These two are,
\begin{itemize}
\item $S$ operation $\left( a_s = d_s = 0, b_s = -c_s = \mathbbm{1}\right)$: This interchanges the electric and magnetic degrees of freedom, which is also an operation quite specific to $3+1$ bulk dimensions.
\item $T$ operation $\left( a_s=b_s = d_s = \mathbbm{1}, c_s =0 \right)$: This results in an additional Chern--Simons term $A \wedge dA$ on the boundary and can be interpreted as adding a quantum of magnetic flux to the electric charge.
\end{itemize}
The two operations do not commute with each other and can be identified with the generators of $SL(2,\Z)$.
Therefore, despite the fact that $Q$ in \eqref{eq:anyon-trafo} can in principle be chosen as an arbitrary $GL(6,\R)$ transformation, we will restrict to the $SL(2,\Z)$ subgroup in the remainder of this paper, which implies the following constraints,
\be
 a_s,\ b_s,\ c_s,\ d_s\in \mathbb{Z} \ \ , \ \ a_s d_s - b_s c_s = 1 \ .
\ee
In this context, the transformed system describes particles carrying $c_s/d_s$ units of original magnetic flux for every unit of original charge.
These particles are precisely the holographic anyons we wish to study.

It is somewhat intricate to work out how the boundary to boundary Green's function transforms under \eqref{eq:anyon-trafo} in general; however, it is relatively straightforward to work it out for $S$ and $T$ operations.
For this purpose, consider plugging in a general transformation into \eqref{eq:S_star} and writing out explicitly,
\be
S^{\ast}
	=	S + \int_{\partial M} \left[
			b_s c_s  \left(A \wedge \star \mathcal{J} - \eta \wedge \star \mathcal{B} \right)
			+ b_s d_s A \wedge \star \mathcal{B}
			+ a_s c_s \eta \wedge \star \mathcal{J} \right]  \, . \label{eq:S_star_II}
\ee
For a $T$ operation, as mentioned above, this simply adds a Chern--Simons term,
\be
- \frac{1}{2 \pi}\int_{\partial M} A \wedge dA
= \int_{\partial M} A \wedge \star \mathcal{B}
\, .
\label{eq:T_trafo_bdrytem}
\ee
The boundary to boundary Green's function is obtained by taking the second derivative of the on-shell action with respect to $A^{\ast}=A$.
`On-shell' in this context means that the equations of motion are solved with some boundary conditions in the bulk -- usually regularity in the origin or in/outgoing conditions at the horizon.
This induces a functional dependence between certain values on the boundary, the details of which are, of course, very dependent on the bulk physics and can not be written down in explicit form, in general.
However, assuming this relation has been worked out for one particular model, from \eqref{eq:T_trafo_bdrytem} immediately follows for the Green's function in the anyonized system,
\be
G^{\ast}_{T} = G + \frac{\partial \mathcal{B}}{\partial A}
\, .
\label{eq:T_trafo_G}
\ee
Working out such a relation for an $S$ operation is a bit more challenging. This is due to the fact that besides interchanging electric and magnetic currents,  it also interchanges bulk fields with their canonical momenta whose explicit form, naturally, depends on the Lagrange density in the bulk.
Nevertheless, in a wide range of cases, \emph{e.g.}, having electromagnetic duality in the bulk or the absence of sources that reach all the way out to the boundary, it can be considered as given that for an $S$-transform on-shell, asymptotically,
\bea
{\star}d A = {\star}d \eta^{\ast} = - 2\pi\mathcal{B} = - 2\pi\mathcal{J}^{\ast} \, , & &
{\star}d A^{\ast} = -{\star}d \eta = - 2\pi\mathcal{B}^{\ast} =  2\pi\mathcal{J} \, .
\eea
Thus, if $G$ is the Green's function of the original action with $\mathcal{J} = G A$, then the following relations are valid,
\be
G^{\ast}_S = - \frac{1}{(2\pi)^2} {\star}d G^{-1}{\star}d \, , \quad
G = - \frac{1}{(2\pi)^2} {\star}d \left(G^{\ast}_S\right)^{-1}{\star}d \, .
\label{eq:S_trafo_G}
\ee
Since any element in $SL(2,\Z)$ is generated by $S$ and $T$, the way the Green's function transforms in full generality can be worked out by successively applying \eqref{eq:T_trafo_G} and \eqref{eq:S_trafo_G}.
A convenient way to summarize the generic result is to define $\mathcal{G}^{-1} := -\frac{\star d}{2 \pi} G^{-1}$, in which case the Green's functions can be related via a modular transform,
\bea
\mathcal{G}^\ast 
&=&	\left( a_s\,\mathcal{G} + b_s \right) \cdot \left( c_s\,\mathcal{G} + d_s \right)^{-1}	\, .
\eea

\section{Application}
\label{sec:application}

In the previous section~\ref{sec:anyonizing} we introduced the general outline of our prescription to obtain a holographic dual for a dense system of anyons in the presence of electric and magnetic fields.
Building up on that, we will now discuss in more detail how a given action transforms under a generic $SL(2,\mathbb{Z})$ mapping that will change the charge carriers into anyons and study its effects on thermodynamics and transport coefficients.
For illustrative purposes, we have chosen to partially focus on a dyonic black brane residing in asymptotically $AdS_4$ spacetime, as this provides a rather instructive setup where all computations can be done entirely analytically.\footnote{For other interesting work relating dyonic black branes to composite fermions see~\cite{Bak:2009kz}.}
Nevertheless, we wish to emphasize that the transformation laws we derive, in particular in section~\ref{sec:transport}, are valid in a more general setting as well.

As an additional remark, for the example of a dyonic black brane we also note that a general transformation with parameters $a_s, b_s, c_s$, and $d_s$ can be neatly expressed in terms of physical quantities, \emph{i.e.}, the filling fraction $\nu$,\footnote{Recall that the Hall conductivity is generally quantized, {\emph{i.e.}}, $
\sigma_{xy}= \frac{e^2}{2 \pi \hbar} \nu$, with $\nu$ being the filling fraction -- broadly speaking the ratio of electric charge to magnetic flux or the extent to which Landau levels are filled -- which either takes integer values (integer QHE) or very specific rational values (FQHE).}
the spontaneous magnetization $m_0$, and $\xi$, which is related to the magnetic susceptibility $\chi_{m}$ via $\xi^2 = -q \chi_{m}$.
The relation can be summarized as follows,
\bea
\left[\begin{array}{cc}
	a_s	& b_s	\\
	c_s	& d_s  \end{array} \right]
&=& \frac{1 }{\sqrt{2 \pi \xi+ \frac{\nu^2 \xi}{2 \pi}}}
\left[\begin{array}{ccc}
	m_0 + \frac{\nu}{2 \pi} \xi	& & m_0 \nu - 2 \pi \xi	\\
	1	& & \nu   \end{array} \right]
\, .
\eea
Or, alternatively, by inverting these equations,
\be
\nu = \frac{d_s}{c_s} \, , \quad
m_0 = \frac{4 \pi^2 a_s c_s + b_s d_s}{4 \pi^2 c_s^2+ d_s^2} \, , \quad
\xi = \frac{2 \pi}{4 \pi^2 c_s^2 + d_s^2} \, .
\ee

\subsection{The dyonic Reissner--Nordstr\"om black brane}
\label{sec:RN}

Here, we set up some conventions and review certain features of this particular spacetime, which will be used later on.
Let us consider the action for Einstein--Maxwell theory,\footnote{For more detail on this model in a holographic setup see, \emph{e.g.},~\cite{Hartnoll:2007ai,Zingg:2011cw}.}
\be
S = - \frac{1}{4 \kappa^2}\int_M \left(R - 2 \Lambda\right) v
	+ \frac{1}{2}\int_M F \wedge *F
	- \frac{1}{2 \kappa^2}\int_{\partial M} K w
\, ,
\ee
where $v = n \wedge w$ with outward normal $n$.
The last term is the usual Gibbons-Hawking term containing the extrinsic curvature $K_{\mu\nu}=\nabla_{(\mu} n_{\nu)}$ on the boundary $\partial M$.
The presence of this term ensures that no additional constraints on derivatives of the metric functions are necessary on the boundary when the action is varied.
Also note that the electromagnetic coupling has been normalized to unity in order to streamline the $SL(2,\mathbb{Z})$ transformation.

Because we are interested in an asymptotically AdS geometry, with cosmological constant $\Lambda=-\frac{3}{L^2}$, this needs to be supplemented with a series of counterterms if the action is to remains finite when evaluated on-shell.
In $(3+1)$ dimensions, this is accomplished by adding a boundary cosmological constant and curvature term,
\be
S_{ct} = \frac{L}{4 \kappa^2}\int_{\partial M} \left( \frac{4}{L^2} + \mathcal{R} \right) w
\, .
\ee
The equations of motion are solved by a dyonic Reissner--Nordstr\"om black brane with metric
\be
\frac{ds^2}{L^2}
	= -f(r) dt^2 + r^2\left(dx^2 + dy^2\right) + \frac{dr^2}{f(r)}	\, ,
\label{eq:metric}
\ee
where the blackening factor reads,
\be
f(r) = r^2 - \frac{1+q_{H}^2+h_{H}^2}{r} + \frac{q_{H}^2+h_{H}^2}{r^2} \ .
\ee
This describes a dyonic black brane where $q_H$ and $h_H$ parametrize charge and magnetic field, respectively.
The corresponding gauge potential and field strength,
\bea
A = \frac{L}{\kappa}\left( q_{H} \frac{r-1}{r}\,dt + h_{H} x \, dy \right)	\, , \quad
F = \frac{L}{\kappa}\left( \frac{q_{H}}{r^2} dr \wedge dt + h_{H} dx \wedge dy \right)
\, .
\eea
After contracting with the normal $n=\frac{dr}{\sqrt{f}}$, the resulting electric and magnetic currents read,
\be 
\mathcal{J} = \frac{q_{H}}{\kappa} \frac{\sqrt{f}\,dt}{r^2}	\, ,\quad
\mathcal{B} = \frac{1}{2\pi} \frac{h_{H}}{\kappa} \frac{\sqrt{f}\,dt}{r^2}
\, .
\label{eq:currents_onshell}
\ee
They are of exactly the same form, reflecting the electromagnetic duality in this setup.
Furthermore, this also makes manifest that the transformation \eqref{eq:anyon-trafo} of the electric and magnetic currents in the present model will essentially result in a mixing of $q_H$ and $h_H$, the electric and magnetic charge at the horizon.
Finally, the component of the magnetic potential that gives us the magnetization in the dual theory can be obtained via integrating $*F$ along the radial direction,
\be
\eta = -2\pi\frac{h_{H}}{\kappa} \frac{r-1}{r}\,dt
\, .
\ee

\subsection{Thermodynamics}
\label{sec:thermo}
Our goal is to describe anyons in the grand canonical ensemble, with natural variables of temperature $T$, chemical potential $\mu$, and magnetic field strength $b$, and corresponding conjugate variables of entropy $s$, charge density $q$, and magnetization $m$.
Thus, we require a family of solutions that has one additional parameter, besides $q_{H}$ and $b_{H}$.
This can easily be constructed by generating solutions with arbitrary horizon radius $r_{H}$ based on \eqref{eq:metric}, via a rescaling of coordinates.
Basic thermodynamic quantities in the boundary theory can then be expressed as follows,
\be
\mu = \frac{q_{H}}{r_{H} \kappa L}	\, , \quad
{q} = \frac{q_{H}}{r_{H}^2 \kappa}	\, , \quad
{m} = -\frac{2\pi h_{H}}{r_{H} \kappa L}	\, , \quad
{b} = \frac{h_{H}}{2 \pi r_{H}^2 \kappa}	\, ,
\nn\ee
\be
{s} = \frac{\pi}{\kappa^2 r_{H}^2}	\, , \quad
{T} = \frac{3-q_{H}^2-h_{H}^2}{4 \pi r_{H} L}	\, . \qquad
\ee
Furthermore (internal) energy, free energy, and Helmholtz free energy density are
\be
{u} = \frac{1 + q_{H}^2 + h_{H}^2}{2 r_{H}^3 \kappa^2 L}	\, , \quad
\Omega = \frac{3 h_{H}^2-q_{H}^2-1}{4 r_{H}^3 \kappa^2 L}	\, , \quad
{a} = \Omega + \mu {q} = \frac{3 h_{H}^2+3 q_{H}^2-1}{4 r_{H}^3 \kappa^2 L}	\, .
\ee
Subsequently, dimensionful constants are omitted for brevity, but can easily be reinstated via dimensional analysis, if needed.
The constants at the horizon are related to the physical quantities at the boundary, 
\begin{eqnarray}
r_{H} &=& \frac{\xi}{2 \pi \left(q- m_0 b\right)} \ , \\
q_{H} &=& \frac{\xi^2 \left( q  \nu  - m_0 b \nu + 2 \pi b \xi\right)}{\left(2 \pi \right)^{3/2} \left(q-m_0 b\right)^2 \sqrt{\left(4 \pi^2+\nu^2\right)\xi}} \ , \\
h_{H} &=& \frac{\xi^2 \left( 2 \pi \left(m_0 b - q\right)  + \nu b \xi \right)}{\left(2 \pi \right)^{3/2} \left(q-m_0 b\right)^2 \sqrt{\left(4 \pi^2+\nu^2\right)\xi}} \ .
\end{eqnarray}
The thermodynamic quantities obey the first law of thermodynamics,
\bea
d \Omega &=&
	- {s}\,d {T}
	- {q}\,d \mu
	- {m}\,d {b}	 \ ,	\\
d {a} &=&
	- {s}\,d {T}
	+ \mu \, d {q}
	- {m}\,d {b}	\, .
\eea
Following the prescription laid out in section~\ref{sec:anyonizing}, we proceed with anyonizing this ensemble.
First, the transformation given in \eqref{eq:anyon-trafo} acts on the boundary quantities as follows,
\bea
\left(
\begin{array}{c}
	{q}_{\ast}	\\
	{b}_{\ast}
\end{array} \right)
= Q \cdot \left(
\begin{array}{c}
	{q}	\\
	{b}
\end{array} \right)	\, , \quad
\left(
\begin{array}{c}
	\mu_{\ast}	\\
	-{m}_{\ast}
\end{array} \right)
= Q^{-T} \cdot \left(
\begin{array}{c}
	\mu	\\
	-{m}
\end{array} \right)
\, .
\label{eq:qbtrafo}
\eea
Then, consider ${a}_{\ast}[{T},{q}_{\ast},{b}_{\ast}]  = {a}[{T},Q^{-1}\cdot ({q}_{\ast},{b}_{\ast})]$ such that, by construction, 
\be
d {a}_{\ast} =
	- {s}\,d {T}
	+ \mu_{\ast} \, d {q}_{\ast}
	- {m}_{\ast}\,d {b}_{\ast}	\, .
\ee
The grand potential in the new ensemble is found via a Legendre transform,
\be
\Omega_{\ast}[{T},\mu_{\ast},{b}_{\ast}] = {a}_{\ast} - \mu_{\ast} {q}_{\ast}	\, .
\ee
This can also be parametrized via values at the horizon,
\be
\Omega_{\ast} = \frac{       (a_s d_s  + 3 b_s c_s)  \pi q_{H}^2
					-(3 a_s d_s + b_s c_s)  \pi h_{H}^2
					+ 2 ( b_s d_s -4 \pi^2 a_s c_s ) h_{H} q_{H}
					+\pi}{4 \pi r_{H}^3} \ .\label{eq:Omega}
\ee

\subsection{Transport properties}
\label{sec:transport}
In order to be able to extract some relevant physical properties of the anyonic fluid in the boundary theory, we will introduce several transport coefficients and perform the anyonization
as discussed above.
We will start by working out the conductivities of the anyonic fluid. The results that we will obtain reproduce those of~\cite{Brattan:2013wya} obtained via a different approach.
Conductivities are related to the Green's function via a Kubo formula,
\be
\sigma_{ij}(\omega)
	= - \frac{\left\langle J_i(\omega) J_j(-\omega)  \right\rangle}{i\,\omega}
	= - \frac{G_{ij}(\omega)}{i\,\omega}
\, .
\ee
The DC conductivities can be obtained in the usual way by taking the zero frequency limit, \emph{i.e.}, $\omega\to 0$.
While finding Green's functions in a given system can be a rather non-trivial task, for conductivities at zero momentum there exists a quite simple, well established formalism.
Firstly, we note that we only need to consider a perturbation of the form,
\be
\delta A = e^{-i\omega t} \left( a_x\, dx + a_y\, dy \right)
\, .
\ee
In this case, we can restrict the Green's function to the spatial components only, so that we can conclude the general form, with the subscript $(H)$ referring to the Hall conductivity,
\bea
G_{ij} = - i\omega \Sigma_{ij}(\omega) \, ,	& &
\Sigma(\omega) = 
\left[\begin{array}{cc}
	\sigma_{(x)}(\omega)	& \sigma_{(H)}(\omega)	\\
	-\sigma_{(H)}(\omega)	& \sigma_{(y)}(\omega)
\end{array} \right]
\, .
\eea
For a $T$ operation, we consider the variation of the magnetic current,
\be
\delta \mathcal{B} = \frac{ i \omega e^{-i\omega t}}{2\pi r} \left( a_y\, dx - a_x\, dy \right)
\, ,
\ee
which gives us the explicit functional dependence on $\delta A$.
Then,  we can apply \eqref{eq:T_trafo_G} directly,
\be \label{eq:condT}
\sigma_{(x,y)}^{\ast} = \sigma_{(x,y)}\ , \qquad  \sigma_{(H)}^{\ast} = \sigma_{(H)} + \frac{1}{2\pi}\ .
\ee
For an $S$ operation, using \eqref{eq:S_trafo_G} yields $G^{\ast} = - \frac{i\omega}{(2\pi)^2} \frac{\Sigma^T}{\det\Sigma}$, and we conclude,
\bea \label{eq:condS}
\sigma_{(x,y)}^{\ast} = \frac{1}{(2\pi)^2} \frac{\sigma_{(x,y)}}{\sigma_{(x)}\sigma_{(y)}+\sigma^2_{(H)}} \ , & &
\sigma_{(H)}^{\ast} = - \frac{1}{(2\pi)^2} \frac{\sigma_{(H)}}{\sigma_{(x)}\sigma_{(y)}+\sigma^2_{(H)}}.\qquad
\eea
In the case of spatial homogeneity, \emph{i.e.}, $\sigma_{(x)}=\sigma_{(y)}$, the results (\ref{eq:condT}) and (\ref{eq:condS}) are in agreement with~\cite{Brattan:2013wya}. 
We can also explicitly apply these transformations to our example from section~\ref{sec:RN}, as the DC conductivities for a dyonic black brane were computed in~\cite{Hartnoll:2007ai}.
With $\sigma^{DC}_{(x,y)} = 0$ and $\sigma^{DC}_{(H)}=\frac{\nu}{2\pi}$ we find,
\bea
 T \ & : & \ \ \sigma_{(x,y)}^{DC,\ast} = 0  \ , \qquad \sigma_{(H)}^{DC,\ast} = \frac{\nu + 1}{2\pi} \ , \\
 S \ & : & \ \ \sigma_{(x,y)}^{DC,\ast} = 0 \ , \qquad \sigma_{(H)}^{DC,\ast} = - \frac{1}{2 \pi \nu} \ .
\eea
The results are in one-to-one correspondence with the generic recipe. Clearly, a $T$ operation corresponds to adding a Chern-Simons term on the boundary, since the corresponding coefficient $\nu\to \nu+1$ is shifted by unity. Moreover, for the $S$ operation the expectation is that the roles of electric and magnetic degrees of freedom be exchanged. This is indeed the case as $\nu \to -1/\nu$. The generic combination of these transformations is
\bea
 \nu \to \frac{a_s\nu+b_s}{c_s\nu + d_s} \, .
\label{eq:DCmodtrans}
\eea
Therefore, it can be concluded that $SL(2,\mathbb{Z})$ acts on $\nu$ as a standard modular transformation.
As a final remark, defining ${{\hat{\Sigma}_i}}^{\,\,k} := 2\pi \Sigma_{ij}\varepsilon^{jk}$ for brevity, (\ref{eq:condT}) and (\ref{eq:condS}) can be summarized as the corresponding special cases of the general modular transform
\bea
\hat{\Sigma} \to 
\left( a_s \hat{\Sigma} + b_s \right) \cdot
\left( c_s \hat{\Sigma} + d_s \right)^{-1}
\, .
\label{eq:DCmodtrans2}
\eea
Susceptibilities, describing the response of the system when applied fields change, are encoded in the Hessian of the grand potential,
\bea
\frac{\partial (s,q,m)}{\partial (T,\mu,b)}
= - \mathrm{H}\Omega (T,\mu,b)
=: \left[\begin{array}{ccc}
	c/T	 		& \lambda_q	& \lambda_m	\\
	\lambda_q	& \chi_{q}	&\chi_{w}	\\
	\lambda_m	& \chi_{w}	& \chi_{m}
\end{array} \right] \, .
\eea
When temperature is varied, the specific heat capacity $c$ describes the change in heat.
Similarly, $\lambda_q$ and $\lambda_m$ encode the corresponding change in charge density and magnetization.
Due to the Maxwell relations, the latter, equivalently, also describes the response of the entropy when chemical potential or magnetic field strength are varied.
The remaining components constitute the electromagnetic susceptibility tensor, with the diagonal elements are often referred to as electric and magnetic susceptibility.
Since \eqref{eq:qbtrafo} encodes the entire functional dependence of the transformed fields in the anyonic ensemble, it is straightforward to calculate how the susceptibilities transform, namely
\bea
\frac{\partial (s,q^{\ast},m^{\ast})}{\partial (T,\mu^{\ast},b^{\ast})}
= - \mathrm{H}\Omega^{\ast} (T,\mu^{\ast},b^{\ast})
= \left[ B - A \cdot \mathrm{H}\Omega (T,\mu,b) \right] \cdot P^{-1}
\, ,
\label{eq:susctrafo}
\eea
where we defined the matrices,
\bea
A = \left[
\begin{array}{ccc}
	1\,	& 					& \hspace{4mm}				\\
	\hspace{4mm}	& a_s	&		\\
		& \hspace{4mm}		& a_s
\end{array} \right] \, , \quad
B = \left[
\begin{array}{ccc}
	0	 			& 					& \hspace{4mm}				\\
		& \hspace{4mm}	& b_s		\\
	\hspace{4mm}	& b_s		& 
\end{array} \right] \, , \quad
P = \left[
\begin{array}{ccc}
	1	 			& 0					& 0				\\
	c_s \lambda_m\;	& c_s\chi_{w}+d_s\;	& c_s\chi_{m}		\\
	c_s \lambda_q	& c_s \chi_{q}		& c_s\chi_{w}+d_s
\end{array} \right] \, .\qquad
\eea
For example, the heat capacity transforms as
\bea
{c}^{\ast} =
	c + \frac{c_s T \left[
			c_s \chi_q \lambda_m^2
			+ c_s \chi_{m} \lambda_q^2
			+ 2 \left( c_s \chi_{w} + d_s \right) \lambda_q \lambda_m
		\right]}{\left( c_s \chi_{w} + d_s \right)^2 - c_s^2 \chi_q \chi_{m}}
\, .
\eea
For $c_s \neq 0$ this will generally mix with the other susceptibilities.
On the other hand, for the electromagnetic susceptibility tensor,
\bea
	\chi_q^\ast
	& = & \frac{\chi_q}{\left( c_s \chi_{w} + d_s \right)^2 - c_s^2 \chi_q \chi_{m}}	\ ,
 \label{eq:qsus}	\\
	\chi_{m}^\ast
	& = & \frac{\chi_{m}}{\left( c_s \chi_{w} + d_s \right)^2 - c_s^2 \chi_q \chi_{m}}	\ ,
 \label{eq:bsus}	\\
	\chi_{w}^\ast
	& = & \frac{\left( a_s \chi_{w} + b_s \right)\left( c_s \chi_{w} + d_s \right) - a_s c_s \chi_q \chi_{m}}{\left( c_s \chi_{w} + d_s \right)^2 - c_s^2 \chi_q \chi_{m}}
\, . \label{eq:wesus}
\eea
This forms a closed set of transformations that does not mix with the thermal responses.
The electric susceptibility of an $ST^K$-transformed anyon fluid was also discussed in~\cite{Brattan:2014moa}, and we see agreement with \eqref{eq:qsus} in that case. 

The transformation rules (\ref{eq:susctrafo}) can also easily be confirmed in the dyonic black brane background, where we can explicitly express all susceptibilities in terms of horizon data,
\bea
	{c}^{\ast}
		& = & \frac{2 \pi (d_s^2+4 \pi c_s^2)(3-q_{H}^2-h_{H}^2)}{8 \pi c_s d_s q_{H} h_{H} + d_s^2 (3+q_{H}^2+3h_{H}^2) + 4\pi^2 c_s^2 (3+3 q_{H}^2+h_{H}^2)}	\ ,
 \label{eq:c_hor}	\\
	\lambda_q^{\ast}
		& = & \frac{4 \pi (2 \pi c_s h_{H}-d_s q_{H})}{8 \pi c_s d_s q_{H} h_{H} + d_s^2 (3+q_{H}^2+3h_{H}^2) + 4\pi^2 c_s^2 (3+3 q_{H}^2+h_{H}^2)}  \ ,
 \label{eq:laq_hor}	\\
	\lambda_m^{\ast}
		& = & \frac{8 \pi^2 (2\pi c_s q_{H} + d_s h_{H})}{8 \pi c_s d_s q_{H} h_{H} + d_s^2 (3+q_{H}^2+3h_{H}^2) + 4\pi^2 c_s^2 (3+3 q_{H}^2+h_{H}^2)}	\ ,
 \label{eq:lam_hor}	\\
	\chi_q^\ast
	& = & \frac{3 (1+q_{H}^2+h_{H}^2)}{r_{H} \left[8 \pi c_s d_s q_{H} h_{H} + d_s^2 (3+q_{H}^2+3h_{H}^2) + 4\pi^2 c_s^2 (3+3 q_{H}^2+h_{H}^2) \right]}	\ ,
 \label{eq:qsus_hor}	\\
	\chi_{m}^\ast
	& = & -\frac{4\pi^2 r_H (3+h_{H}^2+q_{H}^2)}{8 \pi c_s d_s q_{H} h_{H} + d_s^2 (3+q_{H}^2+3h_{H}^2) + 4\pi^2 c_s^2 (3+3 q_{H}^2+h_{H}^2)}	\ ,
 \label{eq:bsus_hor}	\\
	\chi_{w}^\ast
	& = & \frac{4\pi a_s \left[ d_s q_{H} h_{H} + \pi c_s (3+3q_{H}^2+h_{H}^2) \right] + b_s \left[ 4 \pi c_s q_{H} h_{H} + d_s (3+q_{H}^2+3h_{H}^2) \right]}{8 \pi c_s d_s q_{H} h_{H} + d_s^2 (3+q_{H}^2+3h_{H}^2) + 4\pi^2 c_s^2 (3+3 q_{H}^2+h_{H}^2)}		
\, . \qquad\quad \label{eq:wesus_hor}
\eea
We wish to emphasize that we have not fixed a specific $SL(2,\mathbb{Z}$) transformation and thus the parameters $a_s, b_s, c_s$, and $d_s$ appear explicitly in the transport coefficients. Rather than picking a judicious transformation, we wish to advocate the point of view that the measurable quantities of the anyon fluid
will dictate the free parameters $a_s, b_s, c_s$, and $d_s$. The remaining properties of the anyon fluid are then predicted by the dual black brane model. In particular, this does not only apply to static quantities as studied in this paper, but also to dynamic properties, such as the spectrum of quasinormal modes.

\subsection{Low density / high temperature expansion}

To conclude the investigation of the anyonic fluid as described by the dyonic black brane model, we study the equation of state. After all, this was one of the main motivations for this work. As mentioned in the introduction, little is known about the equation of state for an multi particle anyon system using weakly coupled, perturbative methods. On the other hand, using the holographic approach described in this note, we can almost trivially infer the pressure of an anyonic fluid. Recalling that the pressure $P_* = -\Omega_*$ in the grand canonical ensemble, it was already written in its exact form in (\ref{eq:Omega}). However, it is interesting and instructive to expand this expression in various limits.

There are many different ways of expanding the pressure. However, some remarks are in order: Since the anyonic matter we are studying is massless and the bulk solution is asymptotically $AdS$, we expect the boundary gauge theory to be conformal. This means that it is possible to immediately scale out one of the parameters or, equivalently, to simply set $\mu_{\ast}=1$.
Thus, all expressions can, in principle, be given as functions of
${q}_{\ast}$ and ${b}_{\ast}$. Unfortunately, these expressions are rather cumbersome in general. For clarity, we simplify to the case of vanishing magnetic field, $b_\ast = 0$, and we find 
\be
\frac{\Omega_{\ast}}{{T}}	=
	\frac{\pi \left(4 \pi^2 c_s^2+d_s^2\right)^2 \left(1+  \left(4 \pi^2 c_s^2+d_s^2\right)^3 {q}_{\ast}^2  \right) {q}_{\ast}^2}{1-3 \left(4 \pi^2 c_s^2+d_s^2\right)^3 {q}_{\ast}^2 } \ .
\ee
It may appear as if there was a pole in this expression, but it should be kept in mind that ${q}_{\ast}$ is not actually an intrinsic variable, ${T}$ is.
And with $\mu_{\ast}$ and ${b}_{\ast}$ fixed as above,
\be
{T}	= \frac{3 \left(4 \pi^2 c_s^2+d_s^2\right)^3 {q}_{\ast}^2 -1}{4 \pi \left(4 \pi^2 c_s^2+d_s^2\right)^2 {q}_{\ast}}	\, .
\ee
Thus, as the temperature is required to remain non-negative,\footnote{Values with ${q}_{\ast}<0$ would imply $r_{H}<0$.} 
\be\label{eq:Tbound}
 {q}_{\ast} \geq \frac{1}{\sqrt{3 \left(4 \pi^2 c_s^2+d_s^2\right)^3}} \ .
\ee
Hence, an expansion for small ${q}_{\ast}$ only works in a scaling limit where $4 \pi^2 c_s^2+d_s^2 \to \infty$.


Let us now generalize to the case of nonzero magnetic field $b_\ast\ne 0$. To be able to compare the present results for the pressure with the results for an anyonic fluid in the perturbative, weakly coupled regime, \emph{i.e.}, using a virial expansion, it is instructive to expand the normalized transformed free energy $\Omega_{\ast}/{T}^3$, and hence the pressure, in terms of small ${q}_{\ast}/{T}^2$. After a straightforward, albeit lengthy, computation,
\bea
\left.\frac{P_{\ast}}{{T}^3}\right|_{b_{\ast}} &=& \frac{16 \pi^3}{27} + \frac{3 }{4 \xi} \left(\frac{{q}_{\ast}}{{T}^2}\right)^2  -\frac{243 \left(\xi^3 +16 \pi^3 b_{\ast}^2 \left(m_0^2+\xi^2\right)\right)}{ 1024 \pi^3 \xi^5} \left(\frac{{q}_{\ast}}{{T}^2}\right)^4 \nonumber \\\nonumber
&& +\frac{729 \, b_{\ast} m_0 \left(\xi^3 +24 \pi^3 b_{\ast}^2 \left(m_0^2+\xi^2\right)\right)}{ 512 \pi^3 \xi^7} \left(\frac{{q}_{\ast}}{{T}^2}\right)^5\\ && +\mathcal{O}\left(\frac{{q}_{\ast}}{{T}^2}\right)^6 
\, .
\eea
This expression is valid at fixed magnetic field $b_{\ast}$.
For some applications, it may be more convenient to work at fixed ratio $b_{\ast}/q_{\ast}$, yielding the following expansion
\bea
& &\left.\frac{P_{\ast}}{{T}^3}\right|_{b_{\ast}/q_{\ast}} =  \frac{16 \pi^3}{27} + \frac{3\left(\nu^2-4 \pi^2 \left(m_0^2+\xi^2\right)\right)}{4 \nu^2 \xi}\left(\frac{{q}_{\ast}}{{T}^2}\right)^2\nonumber \\ && \qquad + \frac{81\left(4 \pi  \nu  m_0-3 \nu^2+4 \pi^2 \left(m_0^2+\xi^2\right)\right)\left(- 4 \pi  \nu m_0+ \nu ^2+4 \pi ^2 \left(m_0^2+\xi^2 \right) \right)}{1024 \pi^3 \nu^4 \xi^2}\left(\frac{{q}_{\ast}}{{T}^2}\right)^4 \nonumber\\
&& \qquad  +\mathcal{O}\left(\frac{{q}_{\ast}}{{T}^2}\right)^6 .
\eea
A  few remarks are in order. Firstly, for these expansions to be valid, we have to ensure that  the temperature bound \eqref{eq:Tbound}, or an analog thereof at $b_{\ast} \neq 0$, is maintained. In fact, this bound is automatically satisfied, keeping in mind that both expansions can be thought of as the limit $q_{\ast}$ fixed and $T\to \infty$. Secondly, and more importantly, notice that both of the expansions only contain even powers in $q_\ast$, at least at $b_\ast =0$, and in particular no linear term. As an immediate consequence, the equation of state for strongly correlated anyons is drastically different from that obtained using a perturbative expansion, where all virial coefficients are believed to be non-zero. It is tempting to conjecture that strongly correlated anyons, even in more refined holographic models or other types of models in the strongly coupled regime, will behave very differently and resist a naive extrapolation to the ideal gas limit. Future experiments will decide the veracity of this conjecture.

\section{Discussion}\label{sec:discussion}

We conclude this paper with a summary of the advantages of using the formalism as described above.
The Helmholtz action provides a method to treat the currents $\mathcal{J}$ and $\mathcal{B}$ on equal footing, compared to the original action, where the degrees of freedom are $A$ and $\mathcal{B}$, \emph{i.e.}, a 'mix' of a potential and a current. In particular, comparing to \eqref{eq:S_star_II}, this makes the anyonization procedure less contrived and more transparent.
A schematic of the method developed in this paper is summarized in figure~\ref{fig:actiondiagram}.
Additionally, note that the way the auxiliary magnetic potential $\eta$ was introduced does not rely on the bulk equations of motion. In principle, this allows to formulate many details of the procedure of anyonization off-shell.

We then proceeded to study a strongly correlated anyon gas at high temperature. This system is obtained as the gravity dual of a dyonic black brane in $AdS_4$ via an $SL(2,\mathbb{Z})$ electromagnetic transformation. Special emphasis was placed on establishing the invariance of the Helmholtz potential under this transformation. This allowed us to explicitly determine the equation of state of the anyonic fluid at finite density and magnetic fields. The resulting equation of state differs significantly from that of an ideal gas and, amongst other things, the simplifying assumption of (strong) two-body interactions is questionable.
Indeed, when the equation of state is expanded in powers of the charge density, all odd-power coefficients vanish identically, irrespective of the statistics of the underlying charged degrees of freedom.

Our work provides an important step towards a more realistic holographic dual of a strongly correlated anyon fluid.
The method we worked out in section~\ref{sec:anyonizing} can easily be applied to a multitude of cases describing different physics in the bulk -- we only made mild assumptions on the asymptotic behavior of electromagnetism.
An interesting generalization of the dyonic black brane model from section~\ref{sec:application} would be to incorporate fundamental matter, \emph{e.g.}, along the lines of~\cite{Hartnoll:2010ik,Puletti:2010de}.
Such, and other, models have already been utilized when studying the effects of magnetic fields, including quantum oscillations, in the holographic approach on strongly correlated systems~\cite{Denef:2009yy,Albash:2012ht,Blake:2012tp,Gubankova:2013lca,Puletti:2015gwa}.
An application of our framework to these models is straightforward, and could be used to clarify whether an anyon fluid exhibits magnetic oscillations as well.
This exciting possibility may even be soon realized experimentally with ultracold bosonic atoms in optical lattices~\cite{Strater}.

\vspace{0.8cm}
\noindent
{\bf \large Acknowledgments} \;
The authors are grateful to Jarkko J\"arvel\"a, Gilad Lifshytz, Matthew Lippert, and Alfonso Ramallo for many useful discussions and for helpful comments on a draft of this manuscript. 
N.J. and T.Z. have been supported by the Academy of Finland Grants No.~273545 and No.~1268023. 
M.~I. is funded funded by the FCT fellowship SFRH/BI/52188/2013. The Centro de F\'isica do Porto is partially funded by FCT through the projects PTDC/FIS/099293/2008 and CERN/FP/116358/2010.

\appendix

\section{Dependence on boundary data}\label{app:var}

Let us consider a Lagrangian density $\mathcal{L}[A,dA]$ with a $U(1)$ gauge field $A$ on a manifold $M$.
Since we are focusing on $A$, the dependence on other fields will be suppressed in the following. For notational purposes and later convenience we set $\mathcal{H} = {\ast}_4^{-1}\frac{\partial\mathcal{L}}{\partial\, dA}$, where ${\ast}_4$ is the four-dimensional Hodge duality transformation. The variation of the corresponding on-shell action yields
\bea
\delta S
	=	 \delta \int_M \mathcal{L}[A,dA]
	=	\int_M \delta A \wedge \left[ \frac{\partial \mathcal{L}}{\partial A}
				- d{\ast}_4\mathcal{H}	\right]	
			+ \int_{\partial M} \delta A \wedge \star \mathcal{J}
\, . \label{eq:delta_S}
\eea
The last equality uses the definition of the (electric) current
\be
\mathcal{J}
	:= \star^{-1} \frac{\partial \mathcal{L}}{\partial\, dA} 
	= \star^{-1}{\ast}_4\mathcal{H}
	= \imath_n \mathcal{H}^{\perp}
\, . \label{eq:J}
\ee
Hereby, $\star$ is the (three-dimensional) Hodge dual on $\partial M$ and $\imath_n$ denotes the standard inner product for differential forms with the normal direction $n$.
The term in brackets in \eqref{eq:delta_S} are the equations of motion for $A$, which vanish on-shell.
The second term describes the flux through $\partial M$ and thus boundary conditions for it are necessary to make the variation well-defined. These conditions correspond to the boundary degrees of freedom that determine the radial flow, \emph{i.e.}, the components perpendicular to the boundary. It should be kept in mind, however, that there is also boundary data for the components parallel to $\partial M$.
These are identified with a current due to magnetic flux,
\be
\mathcal{B} := - \frac{1}{2\pi} \star^{-1} dA \bigr|_{\partial M}
\, . \label{eq:B}
\ee
In order to proceed with examining the dependence on those degrees of freedom, introducing a consistent notion of what should be identified as parallel and perpendicular directions in the bulk is necessary.
For the purpose at hand, it is convenient to define this via a homotopy operator $\mathcal{K}$ with the following properties,
\begin{enumerate}
\item[i)] 	$\omega = \mathcal{K}d\omega + d\mathcal{K}\omega$ ,
\item[ii)]	$\imath_n \mathcal{K} = 0$ on $\partial M$ .
\end{enumerate}
Then, the `parallel' component can be defined as the projection onto
$*\mathrm{img}\mathcal{K}d^\perp$.
It would go beyond the scope of this paper to have a comprehensive discussion of necessary conditions and topological obstructions to constructing such homotopy operators. However, it should be noted that in the geometries of interest for holographic applications, including the one considered in the main text, \emph{i.e.}, with a pre-existing foliation into parallel submanifolds, such a homotopy operator can be constructed via integration along the radial direction.
Thus, for all intended practical purposes, the existence of $\mathcal{K}$ is presupposed.
After this minor diversion we evaluate the variation with respect to the field strength,
\be
\delta S
	=	\int_M \delta dA \wedge {\ast}_4 \mathcal{H}
	=	\int_{\partial M} \delta dA \wedge \mathcal{K}{\ast}_4 \mathcal{H}
			+ \int_M \delta dA \wedge \mathcal{K} d{\ast}_4\mathcal{H}			
\, . \label{eq:delta_S_mag_II}
\ee
Restricting $dA$ to the parallel component only, and defining the magnetic potential,
\be
\eta := - 2\pi \mathcal{K}{\ast}_4 \mathcal{H} \bigl|_{\partial M}
\, ,
\label{eq:eta_app}
\ee
the variation with respect to the magnetic current can be written in a quite compact form,
\be
\delta S
	= \int_{\partial M} \delta \mathcal{B} \wedge \star \eta
\, . \label{eq:delta_S_mag_final}
\ee


\begin{thebibliography}{99}

\bibitem{Leinaas:1977fm} 
  J.~M.~Leinaas and J.~Myrheim,
  ``On the theory of identical particles,''
  Nuovo Cim.\ B {\bf 37}, 1 (1977).
  doi:10.1007/BF02727953
 
\bibitem{Rao:1992aj} 
  S.~Rao,
  ``An Anyon primer,''
  hep-th/9209066.
  
\bibitem{Stern:2008}
  A.~Stern,
  ``Anyons and the quantum Hall effect: A pedagogical review,''
  Annals of Physics {\bf 323}, 1 (2008)
  
\bibitem{Chen:1989xs} 
  Y.~H.~Chen, F.~Wilczek, E.~Witten and B.~I.~Halperin,
  ``On Anyon Superconductivity,''
  Int.\ J.\ Mod.\ Phys.\ B {\bf 3}, 1001 (1989).
  doi:10.1142/S0217979289000725
  
\bibitem{Moore:1991ks} 
  G.~W.~Moore and N.~Read,
  ``Nonabelions in the fractional quantum Hall effect,''
  Nucl.\ Phys.\ B {\bf 360}, 362 (1991).
  doi:10.1016/0550-3213(91)90407-O
  
\bibitem{Kitaev:1997wr} 
  A.~Y.~Kitaev,
  ``Fault tolerant quantum computation by anyons,''
  Annals Phys.\  {\bf 303}, 2 (2003)
  doi:10.1016/S0003-4916(02)00018-0
  [quant-ph/9707021].
 
\bibitem{Hosotani:1992jm}
  Y.~Hosotani,
  ``Neutral and charged anyon fluids,''
  Int.\ J.\ Mod.\ Phys.\ B {\bf 7} (1993) 2219
  doi:10.1142/S0217979293002857
  [cond-mat/9302002].
  
  
\bibitem{YiCanright}
 J.~Yi and G.~S.~Canright,
  ``Spontaneous magnetization of anyons with long-range repulsion,''
  Phys.\ Rev.\ B {\bf 47} (1993) 273
  doi:10.1103/PhysRevB.47.273
  
\bibitem{AdS_CFT_reviews} 
 J.~Casalderrey-Solana, H.~Liu, D.~Mateos, K.~Rajagopal and U.~A.~Wiedemann,
  ``Gauge/String Duality, Hot QCD and Heavy Ion Collisions,''
  arXiv:1101.0618 [hep-th];
 J.~McGreevy,
  ``Holographic duality with a view toward many-body physics,''
  Adv.\ High Energy Phys.\  {\bf 2010}, 723105 (2010)
  [arXiv:0909.0518 [hep-th]];
    A.~V.~Ramallo,
  ``Introduction to the AdS/CFT correspondence,''
  Springer Proc.\ Phys.\  {\bf 161} (2015) 411
  [arXiv:1310.4319 [hep-th]].

\bibitem{Witten:2003ya} 
  E.~Witten,
  ``SL(2,Z) action on three-dimensional conformal field theories with Abelian symmetry,''
  In *Shifman, M. (ed.) et al.: From fields to strings, vol. 2* 1173-1200
  [hep-th/0307041].
  
\bibitem{Yee:2004ju} 
  H.~U.~Yee,
  ``A Note on AdS / CFT dual of SL(2,Z) action on 3-D conformal field theories with U(1) symmetry,''
  Phys.\ Lett.\ B {\bf 598}, 139 (2004)
  doi:10.1016/j.physletb.2004.05.082
  [hep-th/0402115].
  
\bibitem{Jokela:2013hta} 
  N.~Jokela, G.~Lifschytz and M.~Lippert,
  ``Holographic anyonic superfluidity,''
  JHEP {\bf 1310}, 014 (2013)
  [arXiv:1307.6336 [hep-th]].

\bibitem{Bayntun:2010nx}
  A.~Bayntun, C.~P.~Burgess, B.~P.~Dolan and S.~S.~Lee,
  ``AdS/QHE: Towards a Holographic Description of Quantum Hall Experiments,''
  New J.\ Phys.\  {\bf 13} (2011) 035012
  doi:10.1088/1367-2630/13/3/035012
  [arXiv:1008.1917 [hep-th]].
 
 \bibitem{Goldstein:2010aw} 
  K.~Goldstein, N.~Iizuka, S.~Kachru, S.~Prakash, S.~P.~Trivedi and A.~Westphal,
  ``Holography of Dyonic Dilaton Black Branes,''
  JHEP {\bf 1010}, 027 (2010)
  doi:10.1007/JHEP10(2010)027
  [arXiv:1007.2490 [hep-th]].
  
\bibitem{Fujita:2012fp}
  M.~Fujita, M.~Kaminski and A.~Karch,
  ``SL(2,Z) Action on AdS/BCFT and Hall Conductivities,''
  JHEP {\bf 1207} (2012) 150
  doi:10.1007/JHEP07(2012)150
  [arXiv:1204.0012 [hep-th]].
  
\bibitem{Lippert:2014jma} 
  M.~Lippert, R.~Meyer and A.~Taliotis,
  ``A holographic model for the fractional quantum Hall effect,''
  JHEP {\bf 1501}, 023 (2015)
  [arXiv:1409.1369 [hep-th]].

\bibitem{Bergman:2010gm}
  O.~Bergman, N.~Jokela, G.~Lifschytz and M.~Lippert,
  ``Quantum Hall Effect in a Holographic Model,''
  JHEP {\bf 1010} (2010) 063
  doi:10.1007/JHEP10(2010)063
  [arXiv:1003.4965 [hep-th]].
  
\bibitem{Jokela:2014wsa} 
  N.~Jokela, G.~Lifschytz and M.~Lippert,
  ``Flowing holographic anyonic superfluid,''
  JHEP {\bf 1410}, 21 (2014)
  doi:10.1007/JHEP10(2014)021
  [arXiv:1407.3794 [hep-th]].
  
\bibitem{Brattan:2013wya} 
  D.~K.~Brattan and G.~Lifschytz,
  ``Holographic plasma and anyonic fluids,''
  JHEP {\bf 1402}, 090 (2014)
  doi:10.1007/JHEP02(2014)090
  [arXiv:1310.2610 [hep-th]].

\bibitem{Brattan:2014moa} 
  D.~K.~Brattan,
  ``A strongly coupled anyon material,''
  JHEP {\bf 1511}, 214 (2015)
  doi:10.1007/JHEP11(2015)214
  [arXiv:1412.1489 [hep-th]].
 
\bibitem{Itsios:2015kja}
  G.~Itsios, N.~Jokela and A.~V.~Ramallo,
  ``Cold holographic matter in the Higgs branch,''
  Phys.\ Lett.\ B {\bf 747} (2015) 229
  [arXiv:1505.02629 [hep-th]].
  
\bibitem{Jokela:2015aha}
  N.~Jokela and A.~V.~Ramallo,
  ``Universal properties of cold holographic matter,''
  Phys.\ Rev.\ D {\bf 92} (2015) 2,  026004
  doi:10.1103/PhysRevD.92.026004
  [arXiv:1503.04327 [hep-th]].
  
\bibitem{Itsios:2016ffv}
  G.~Itsios, N.~Jokela and A.~V.~Ramallo,
  ``Collective excitations of massive flavor branes,''
  arXiv:1602.06106 [hep-th].

\bibitem{Bak:2009kz}
  D.~Bak and S.~J.~Rey,
  ``Composite Fermion Metals from Dyon Black Holes and S-Duality,''
  JHEP {\bf 1009} (2010) 032
  doi:10.1007/JHEP09(2010)032
  [arXiv:0912.0939 [hep-th]].

\bibitem{Hartnoll:2007ai}
  S.~A.~Hartnoll and P.~Kovtun,
  ``Hall conductivity from dyonic black holes,''
  Phys.\ Rev.\ D {\bf 76} (2007) 066001
  doi:10.1103/PhysRevD.76.066001
  [arXiv:0704.1160 [hep-th]].
 
\bibitem{Zingg:2011cw} 
  T.~Zingg,
  ``Thermodynamics of Dyonic Lifshitz Black Holes,''
  JHEP {\bf 1109}, 067 (2011)
  [arXiv:1107.3117 [hep-th]].

\bibitem{Hartnoll:2010ik}
  S.~A.~Hartnoll and P.~Petrov,
  ``Electron star birth: A continuous phase transition at nonzero density,''
  Phys.\ Rev.\ Lett.\  {\bf 106} (2011) 121601
  doi:10.1103/PhysRevLett.106.121601
  [arXiv:1011.6469 [hep-th]].
  
\bibitem{Puletti:2010de}
  V.~G.~M.~Puletti, S.~Nowling, L.~Thorlacius and T.~Zingg,
  ``Holographic metals at finite temperature,''
  JHEP {\bf 1101} (2011) 117
  doi:10.1007/JHEP01(2011)117
  [arXiv:1011.6261 [hep-th]].

\bibitem{Denef:2009yy} 
  F.~Denef, S.~A.~Hartnoll and S.~Sachdev,
  ``Quantum oscillations and black hole ringing,''
  Phys.\ Rev.\ D {\bf 80}, 126016 (2009)
  doi:10.1103/PhysRevD.80.126016
  [arXiv:0908.1788 [hep-th]].

\bibitem{Albash:2012ht} 
  T.~Albash, C.~V.~Johnson and S.~MacDonald,
  ``Holography, Fractionalization and Magnetic Fields,''
  Lect.\ Notes Phys.\  {\bf 871}, 537 (2013)
  [arXiv:1207.1677 [hep-th]].

\bibitem{Blake:2012tp} 
  M.~Blake, S.~Bolognesi, D.~Tong and K.~Wong,
  ``Holographic Dual of the Lowest Landau Level,''
  JHEP {\bf 1212}, 039 (2012)
  doi:10.1007/JHEP12(2012)039
  [arXiv:1208.5771 [hep-th]].
  
\bibitem{Gubankova:2013lca} 
  E.~Gubankova, J.~Brill, M.~Cubrovic, K.~Schalm, P.~Schijven and J.~Zaanen,
  ``Holographic description of strongly correlated electrons in external magnetic fields,''
  Lect.\ Notes Phys.\  {\bf 871}, 555 (2013)
  [arXiv:1304.3835 [hep-th]].

\bibitem{Puletti:2015gwa}
  V.~Giangreco M. Puletti, S.~Nowling, L.~Thorlacius and T.~Zingg,
  ``Magnetic oscillations in a holographic liquid,''
  Phys.\ Rev.\ D {\bf 91} (2015) 8,  086008
  doi:10.1103/PhysRevD.91.086008
  [arXiv:1501.06459 [hep-th]].

\bibitem{Strater}
  C.~Str\"ater, S.~C.~L.~Srivastava and A.~Eckardt
  ``Floquet realization and signatures of one-dimensional anyons in an optical lattice,''
  arXiv:1602.08384.


  \end{thebibliography}
\end{document}